# Exploration of the Gap Between Computer Science Curriculum and Industrial I.T Skills Requirements.

**Azeez Nureni Ayofe**

Department of Maths & Computer Science,

College of Natural and Applied Sciences,

Fountain University, Osogbo,

Osun State, Nigeria.

E-mail address: **nurayhn@yahoo.ca**

**Azeez Raheem Ajetola**

Department of Maths & Computer Science,

College of Natural and Applied Sciences,

Fountain University, Osogbo,

Osun State, Nigeria.

E-mail address: **ajeazeez@yahoo.com**

**ABSTRACT**

This paper sets out to examine the skills gaps between the industrial application of Information Technology and university academic programmes (curriculum). It looks at some of the causes, and considers the probable solutions for bridging the gap between them and suggests the possibilities of exploring a new role for our universities and employers of labor. It also highlights strategies to abolish the misalignment between university and industry. The main concept is to blend the academic rigidity with the industrial relevance.

## 1.0     INTRODUCTION

As the Nigerian industries are rapidly growing in terms of the advancement of science and technology, unprecedented demand for better graduates has been created. However, industry often criticizes that existing university curricula fall short to tackle the practical issues in the industry. For instance, the industry expects the university to train their future employees with the latest technology. Academia is at the centre of developing trends. This is because university lacks a proper academic programme that is suitable for the industries. This causes a gap between universities and industry that needs to be bridged by the universities academics and IT professionals. The industry is continually broadening and the knowledge domain is increasingly becoming complex. The importance and role of developing better curriculum in universities programme is significant in bridging the gap between the changing technology and industry needs for employers. Universities should provide a conducive learning environment and industry oriented curriculum that the business community perceived as meeting their IT requirements. Curricula are expected to be developed with the objective of producing skilled and employable graduates. Ching *et al* (2000) states that employability rests in the knowledge and skills imparted upon them through their education.

This paper therefore sets out to examine the skills gaps between the industrial application of Information Technology and university academic programmes, look at some of the causes, and in considering the probable solutions for bridging the gap between them and suggests the possibilities of exploring a new role for our universities and employers of labor. The two sides, one producing and the other utilizing the work force, need a common ground to operate so that such synergy will result in adequate supply of relevant personnel for all the sectors of the economy.

It is when such a balance is in sight that we may begin to wrap our arms around resolving the issue of unemployment in the society.

## 2.0  UNIVERSITY ACADEMIC PROGRAM AND INDUSTRIAL APPLICATION OF IT

The subject of skills development is not only timely but appropriate in view of the present global socio-economic challenges. The issue of skills gap is particularly topical considering the structural, academic, vocational and planning challenges which are peculiar to us presently. No longer is the world debating on the importance of education as a pre-requisite for social and economic development, and nobody now questions the relationship between high academic attainment and economic rewards that accrue as a result of that attainment. The former President of the United States Bill Clinton once said "We are living in a world where what you earn is a function of what you can learn. (US Dept. of Educ., 1995).

If this world is to move out of the present economic doldrums, then its abundant human resources needed to be deployed effectively and efficiently with the skill of Information Technology based processing to manage other natural resources, in order to attain these developmental goals.

IT Skills in all and every ramification translate into inventions, services, products, ideas, innovations and best practices that drive the wheel of progress and development. From a studied position, the development of any nation depends to a very large extent on the caliber, organization and technological skill of its human resources.

In addition, it is widely held that knowledge, skills, and resourcefulness of people are critical to sustaining economic and social development activities in a knowledge based society. Given the growing global IT networking and the dynamic investment climate in the world, the demand for knowledge workers with high levels of technical and soft skills can only increase. IT knowledge and networking skills is the arrowhead of the modern world of work. All aspect of work is now computerized. Only those who move with the tide will be successful.

However, the gap that exists between what is taught at school and the skills required to perform on a job is so wide that a high percentage of young graduates are said to be unemployable for lack of needed skills that would make them profitable for any employer. This state of affairs has existed in the world especially in Africa for so long that there is urgent need for serious actions to stem the tide and correct the malaise that is robbing the nation of progress in many fields of endeavour.

### 3.0 A TYPICAL SCENARIO

The table1 below shows the statistics of unemployed graduates in Malaysia as obtained in (http://educationmalaysia.blogspot.com/2006/07/70-public-university-graduates-jobless.html), as demonstrated during a seminar in Malaysia on Education in Malaysia.

| Courses / Subjects | Unemployed | % |
|---|---|---|
| Computer Science | 3,942 | 19.5% |
| Business Administration / Management | 3,736 | 18.5% |
| Engineering | 3,096 | 15.3% |
| Accountancy | 1,923 | 9.5% |
| Literature & Social Sciences | 1,283 | 6.3% |
| Pure Science & Applied Sciences | 1,303 | 6.4% |
| Architecture & Building Management | 540 | 2.7% |
| Agriculture, Fisheries & Forestry | 401 | 2.0% |
| Others | 3,993 | 19.8% |
| TOTAL | 20,217 | 100.0% |

**Table 1** shows the statistics of unemployed graduates in Malaysia (source: http://educationmalaysia.blogspot.com/2006/07/70-public-university-graduates-jobless.html)

One of the contributors, Kian Ming, said "I can fully understand "Business Administration" or other management programmes as a degree course that many candidates opt for if they are not qualified for other subjects to study, and hence the high level of unemployability given the weaker pool of students. However, computer science has the highest contributor to the unemployed pool? Isn't that the next wave of growth overtaking the country whereby computer science graduates should be in high demand?"

Another participant in the same seminar, John Lee, also said "The answer as to why the Computer Science faculty seems to be contributing the highest number of unemployed graduates to the market place despite a clear shortage of skilled workers in the industry is fairly obvious.

A survey conducted earlier has indicated that as many as 30% of the unemployed local graduates are computer science and information technology degree holders. These skills are in obvious demand in the country - it is not a mismatch. The clear-cut issue in this case is that many of the local institutions of higher learning, both public and private have failed to offer a sufficiently rigorous education to produce the necessary quality in the workforce which the industry requires"

Most importantly, as highlighted by Chris Chan, chief executive officer of The Media Shoppe in the same seminar, he said:

"… some local ICT graduates lacked fundamental technical skills and only had knowledge of basic software such as Microsoft Office (!)
The problem is largely either the poor ICT curriculum of many of our local universities/colleges that doesn't seem teach anything to our ICT students or these students shouldn't have been taking ICT courses in the first place"

**4.0  WHAT IS A SKILL GAP?**  A skill gap is the shortage in performance. It is the difference between what is required or expected and what we actually get. Put in another way a skill gap is the required performance minus the present performance (Adetokunbo 2009). Hence it is also called the **performance gap**. It could be in the area of any respective field of work.

**Causes of gap between the university degree in Computing and industrial IT skills**

- The Computer Science curriculum is static in nature while its industrial application is dynamic.
- University is not ready to train and retrain its staff to meet up with the dynamic nature of the course because of the financial implication.
- Lukewarm attitude of lecturers to surrender themselves for training and workshops that will expose them to the latest innovations in IT.
- Priority given to research works by the lecturers rather than lectures and workshops which will bring them to limelight on the latest development in IT.
- Lack of facilities to train both the lecturers and the students on the new inventions

**4.1 UNIVERSITY ACADEMIC PROGRAMME:**
This otherwise known as 'Curriculum' refers to course offerings at an educational institution. Decisions about what a school should teach are usually made by school administrators and faculty and governed by University councils.
In relation to Information Technology, it is of the view as it being too theoretical and outdated. The necessary technical attributes and "Know-how" expected of this program is in a depleted state and close to nothing satisfactory to applications in the Industrial realm.

Answers are continuously left un-provided when students (graduates) are faced with the reality question of: "**WHAT CAN YOU DO?**" in the labour market when they are out for any interview.

**4.2  STUDENTS IN PERSPECTIVE**
It is clearly obvious that in university, students study the basics, that is, underline principles, which might not be adequate to develop a professional project for a good client

Students do not know what **a use case** is; they also do not know how to prepare a professional SRS. They equally do not know about the WBS. So how can they learn all these to prepare themselves joining a good satisfying job and work confidently?

They should not think that they know in and out of software development the moment they get a degree certificate from the university.

They must accept the fact that may be they know 10% or they just heard about all these jargons during their student life. They should also educate their parents not to pressurize them just after their graduation rather to co-operate with them to learn and get ready for a right job.

Why so many fresh IT or Computer Science graduates in India could avail the job within 6 months time from the graduation date? They set their mind to join a good training institute at least for 3 months time after the graduation where they learn the technology, communication and exposure on project management. This helps them a lot to approach the big companies for a junior software developer post as they learn the live scenario of a project cycle during their training period (Azeez N.A, 2008). They know how to code (technology), how to document (communication) and how to prepare a release note (bit of project management). This is what any company expects from any IT graduate from day one.

They feel happy recruiting them as they don't have to spend money and time providing training to such a graduate any longer.

### 4.3 CAUSES OF UNIVERSITY- INDUSTRY GAP IN THE AREA OF INFORMATION TECHNOLOGY

Apart from skill obsolescence that occurs over time, there are other factors that cause these gaps. ***A major factor is the changing pattern of working in industries***. The current trends in the world of work such as globalization, commercialization, flexi-hour, deregulation, outsourcing, contract work, homework and freelancing have led to marked changes in industry structure. **New** definition, **new** meaning, and **new** application of knowledge drive all these changes**. New** technological discoveries have given rise to new industries and new structuring of work itself. **New** forms of work structures which are flexible, adaptable, less hierarchical, and multi-skilled and which encourage continuous learning are becoming sources of competitive advantage in industries. International competition for jobs and workers has also intensified, leading to the global talent ***hunt for innovation-driven knowledge workers.***

In addition, ***global organizations are finding themselves ill equipped to compete in the 21st Century because of lack of right skills in fresh graduates that are employed in the labor market.***.

At a time when the global knowledge-based economy places an ever-growing premium on the talent, creativity, and efficiency of the workforce, business leaders talk of a widening gap between the skill their organizations need to grow and the capabilities of their employees. Finding the right candidates to fill a growing list of vacant positions is a number one concern of business leaders today. Research shows that the shifts in workforce demographics affect the availability of labor to fill high-skilled jobs. Ironically, skill gaps result from technological advancements. *Therefore, in reality, organizations will always face some types of skill gaps all the time if the university curriculum does not adjust itself into the computerized economy.*

Lack of proper skills in the university students, re-skilling, poor facilities for IT skills development , lack of planning, lack of coordination, confusion, mismanagement, inefficient application of scarce resources and deficient value orientation and other perfidies has greatly contributed to put our country in a very precarious job deficit.
Information technological Training facilities are few, uncoordinated and untargeted in the higher institutions. Before the current global economic crisis, the jobs deficit was already huge and unwieldy. The situation has now become even more critical.

A respondent in a current research carried out commented upon lack of teaching staff and administrative difficulties in updating the university programmes curriculum for IT education.
Lack of technical expertise, costly IT equipments, costly maintenance and replacement of equipments, have been some of the major impediments.

Another major problem has been the schools' inability to keep abreast of fast changing developments in industry and technology.

It was established earlier that a gap exists between subjects taught and the methods used to teach these subjects, and the academic requirements at higher education institutions.

### 4.4  DIMENSIONS OF SKILL GAP

From the foregoing analysis, it becomes obvious that there will always be a gap between IT skills and the university degree in Computing regardless of the operative economic system. The extent and life span of the gap depends on how fast universities adjust or update their curricular respond to structural changes, and the magnitude, composition and time-lag in government intervention in the labor market. Gaps therefore exist in various forms at the aggregate, sectoral and individual levels.
Underlying this gap is **inadequacy of the educational curricula** which is designed without apparent regards for relevance in application of the industries. Aside from this, there is some lop-sidedness in curricula implementation. Frequent discontinuity in the university program impact directly on the quality of skills supplied. The short-term duration practical exposure of students through the SIWES (**Student Industrial Work Experience Scheme**) is generally ineffective. This is because most higher institutions do not even have proper IT facilities.

### 4.5 CONSEQUENCES OF THE GAP

The persistent existence of skills gap in the IT industries and universities has made dependence on importation of skilled workers with its attendant cost inevitable (Adetokunbo, 2009).

The gap also results in a **waste of human resources** and, therefore, unemployment. For example, banks in many parts of African countries usually purchase software which they use in banks for transactional purposes from China or United States of America. Also, big companies equally do same for smooth running of their day to day activities. This is sequel to lack of reliable IT personnel in many parts of the world to take over the challenge. Inappropriately skilled labor is deprived of participation in the production process. This category of unemployed persons raises the noise level of unemployment which in addition to its economic consequences also threatens social stability of the country.

### 4.6 BRIDGING THE SKILLS GAP

What then can be done to bridge this gap? What kind of education is required in order to prepare our students for work in the industries? What changes need to be implemented in order to make university programmes suitable for a true preparation for work? What kind of program would ensure that students possess the skills necessary to enable them to occupy the jobs currently taken by expatriates?

The answer to this question evidently lies in exposing the students to the high-level cognitive skills that are essential and required by industries. The following are some of the solutions that have been found to produce good results:

- **Study IT Skills program:** This is normally either presented as a stand-alone program or integrated into subjects taught. IT Skills are reinforced, self-learning, lifelong learning, research skills, time management skills, critical thinking skills etc. These components have been seen to be most effective when they are woven into the university curriculum rather than tackling them as stand-alone subjects.

In bridging the gap and/or reforming education, many countries have encountered and addressed this issue by introducing a strong technological component to the university curriculum. This normally comes in many different forms; prevalent among them is offering students courses in IT, work attitude and work ethic, followed by a subsequent placement in industrial and commercial firms, where they get firsthand experience in real work environment.

Successful programs have been implemented in countries like Australia, Canada, United States, and Britain. The success of such programs in these countries is ensured by the existence of a huge industrial sector, which works in partnership with schools. Other countries have opted to establish training centers, which have workshops that give students real work experience. These training centers are normally set up, financed, and managed by the private sector and schools pay fees for their students to use these centers. A successful example of this kind of programs can be seen in the BOCES program in New York State, and the Chicago School-to-Work Program.

- **Information Technology:** This program ensures that the student possesses adequate Knowledge of IT and has the skills required to comfortably use it in his job. Knowledge and skills of IT are two components that have been found to be essential to both groups of students; the one that joins the workforce and the one that opts for higher education.

Like every economic phenomenon, there are both supply and demand sides to issues relating to skills. A major source of supply of skills is the educational system which is defined by the totality of all formal educational institutions providing one form of skills development or the other ranging from the basic, technical colleges, to tertiary institutions comprising the various Universities, Polytechnics, Monotechnics and other specialized institutions providing highly specialized skills.

- ➢ The curricula or training manuals being implemented by these various institutions are developed either wholly by the

relevant coordinating commissions such as the National Universities Commission {NUC} in case of the Universities, the National Board for Technical Education {NBTE} for the Polytechnics, or in conjunction with international agencies such as the ILO. Suffice it to say that the major target of the educational system is to produce skills required by the public and the organized private sector.

- Government and the organized private sector should also put in place arrangements for professional students of tertiary institutions to undergo short-term practical training in their chosen vocations through a **Student Industrial Work Experience Scheme [SIWES]** in Information Technology to enhance their knowledge in the field.

There is an emerging group of skills developers in Information and Communication Technology who can be placed between institutional and private developers. The emergence of this group is in response to developments in the ICT industry. Government should make promotional efforts towards regulating operations in the IT sector to avoid possible lop-sidedness and unhealthy practices that could mar the sector.

- **Appropriate educational curricula**

This must be designed and implemented by our institutions of learning especially the technical colleges, polytechnics, monotechnics, universities and other specialized training institutions. The curricula which must be relevant to the peculiarities of our situation must address most importantly the current industrial demands with the intention of making our university graduates of Computer Science relevant in the IT industry.

- There is need to actively collaborate and involve employers of labor in developing appropriate IT skills to avoid the situation whereby people trained in certain field cannot utilize them while skills needed by employers are non available or grossly inadequate leading either to importation of foreign skills or outright incapacitation of the production process. Employers should be involved in all forms and levels of skills development ranging from curricula design and implementation, product / service research and development, funding, etc. The need to institutionalize **Entrepreneurship Development Programme** {EDP} and vocational training in the educational curricula is also imperative. Happily enough, some institutions have already started this. These subjects should be included among the contents of the compulsory general studies programme of all tertiary institutions' IT curricula.

- **Dialogue between the universities and employers of labor;**

An outline for a framework for fostering the partnership for interaction between university and employer, while some areas of positive interaction between university and employers exist in the forms of training programs, and joint services geared at bridging the skill gaps, what is needed, however is a framework that addresses the chronic skilled shortages in the labor market. This no doubt will entail an integrated strategy.

**4.7 EMPLOYERS' PERSPECTIVE** Employer-university interaction is currently characterized by problem of skills mismatch between what employer want and what university can provide. So the Universities must design a proper programme for the proper identification of employers' skills requirements. For a result-oriented dialogue therefore, and on the part of employers, they should do the following in order to attain maximum benefits that will be accrued in bridging the current gap between the university curriculum of computer science and IT skill requirements in the industry:-

**Educational Reform/Curriculum**:
Educational reform is the most important area in which university can aid in bridging this gap. The rapidly changing needs of employers and the labor market affect curriculum. Adjusting the curriculum to rapidly changing needs of employers and the labor market is therefore very imperative. In framing an innovative curricular relevant to employers' need for IT, universities must factor in the dynamics of modern trends, including ICT, globalization and technological changes. Technology not only has given rise to vast new industries, but the restructuring of work itself. New forms of work structures which are flexible, adaptable, less hierarchical, multi-skilled, and continuous learning are becoming one of the major sources of competitive advantage of enterprises in IT industries.

**ICT literacy:** Literacy in ICT must become an imperative of the educational process and integrated into the curriculum at all levels of studies to match the challenges and opportunities before us. Our objective is to empower every citizen with the IT skills they need for life-long learning, both in the workplace and in private life. Our citizens must have the technical skills, confidence, and flexibility they need to adapt over the course of their lifetimes.

**Industry's needs:** The drivers for adoption of a productive dialogue with university/working partnership with industry on skill development in relation to the curriculum include knowing the industry's skills requirements. What skills make graduate more employable? (These may include, for example, in Computer Science, a programmer is expected to be a mastery of the following programming languages: Java, C++, C, DHTML, Oracle 10g, ASP and CGI as well as C# for him to be relevant and employable in the labour market.

Also, the following categories of professionals are expected to be able to perform the following functions:

1. A trained Software Engineer is expected to know how to create, maintain and modify computer and software programs such as operating systems, communications software, utility programs, compilers and database handlers. They may also be able to evaluate new programming tools and techniques and analyze current software products (http/www./scientist/223113A.htm).
2. Computer engineers are involved in the installation, repair and servicing of computers and associated equipment, or peripherals. They may sometimes be described as information technology (IT) hardware technicians, service engineers or computer systems engineers (http/www./scientist/223113A.htm).
3. A hardware design engineer plans, designs, constructs and maintains the hardware equipment of computers. They may also monitor the development of hardware according to design, and carry out repairs and testing of computer equipment and peripherals (http/www./scientist/223113A.htm).
4. A network/systems engineer designs, installs, analyses and implements computer systems/networks. They may also make sure that the existing network is effective, and work out how it should evolve to meet new requirements of the organization or business (http/www./scientist/223113A.htm).

The question now is, are these categories of professionals in Computing being trained to acquire the above skills? The answer is no, this can be clearly established base on the analysis done above.

These are highly required to be taught in the university but reverse is the case.

**Employability of graduates:** In order to overcome persistent mismatches between graduate qualifications and the needs of the labor market, university programmes should be structured to enhance directly the employability of graduates and to offer broad support to the workforce more generally. ICT Skills are portable if the skills acquired are transferable and can be used productively in different jobs, enterprises, both in the informal and formal economy. Emphasis should be placed on entrepreneurship development to make our graduates well equipped for self employment, innovation and creativeness.

**5.0 CHALLENGES TO UNIVERSITIES**
The implication of many of the processes of globalization, knowledge redefinition, graduate employability etc, is yet to be addressed by most universities. The scale of the challenge should however not be underestimated. Indeed, becoming a market-responsive organization requires a major change in university culture. It implies a strong sense of institutional purpose and redirection through re designing the university academic curriculum the Computer Science graduate relevant in their chosen field.

**Governance, Management and Leadership.**
Universities have historically been run as community of scholars. Governance and management structures were collegial and committee-based, the Senate and the council were representative, and therefore, large. Decision making, was as a result slow and naturally conservative. The emergence of a competitive mass market and global higher education market is bringing this model of governance and management into question. If we are to have a catalytic relationship between university and the global and dynamic world of the industries it is vital for the universities to transform also into more dynamic institutions. In short, improved dialogue between universities and industries will not be readily achieved by top down mechanisms at either the institutional or regional level. There is thus the need for a flexible, responsive and agile organization able to strike a working partnership with others.

## 5.1 BENEFITS TO THE UNIVERSITIES

The US experience at developing the 'knowledge workers' is a good example to demonstrate that universities do play a vital role in driving growth in the modern economy. If this paper is fully read and digested, the universities should get sufficient information on the skills needs of the industries to convince them of the urgency in curriculum refurbishing. Still, there are clear benefits that will accrue to them from this paper.

(1) They will have enhanced role to participate in industrial economic development.
(2) They will have earmarked funding for specific projects and research efforts from more industries.
(3) They will have flexible plans to access research funding through collaboration with industries.
(4) They will have access through enlarged programmes to real world challenges in the workplace; and have the satisfaction of contributing to market place success and growth ideas.
(5) They will have access to modern and sophisticated equipment and facilities in research centers funded by industries jointly or otherwise.
(6) The need will create worthwhile incentives to help recruit, reward and retain research and faculty members and most importantly to train employable graduate.

## 5.2 BENEFITS TO BUSINESS AND INDUSTRIES

Some specific benefits of this paper also lead to acquiring the industrial and the business world advantages. These include the following:

(1) There will be a steady and constant supply of graduate and post-graduate talents, skilled in the needed areas for employment.
(2) A pool of scientists and researchers will be available to undertake regular projects that will keep the industries abreast of innovations and discoveries.
(3) The availability of the latest research and technological inventions in the Nigerian market place would be guaranteed.
(4) Nigerian industrialists and academicians will rub shoulders with their international counterparts in intellectual networking.
(5) The need for constant upgrading of professional knowledge becomes imperative for lecturers, staff and management alike.

The foregoing in summary, underscores the need to build partnerships between universities and industries in Information Technology and research-intensive sectors. Many multinationals have established alliances with academic institutions on specific initiatives covering faculty upgrading, consultancy, internships, curriculum revision workshops, research incubation, etc. aggregating the architects of the new global development in educational sector.

**Bridging the gap: Student efforts.**
➢ In summary, students should find a good training company where they should not spend more money and time but can learn more professionally to augment their degree certificates.

➢ Fresh graduates must think and plan about their career whether to become a Programmer, Business Analyst, Project Manager, Architect or preparing the career in Sales & Marketing before graduation.

They must think about the career path – how to achieve their career goal within a certain number of years

## 6.0 RECOMMENDATIONS AND CONCLUSION

Whatever the format of education that will be agreed upon, the present researcher believes that there are some important parameters that need to be established. These parameters call for a paradigm shift from "Instruction" to "Learning" and from the "Sellers' market" to the "buyers' market" (UNESCO, 2001). This shift also calls for a solid and sustained collaboration between education and the community. New partnerships would therefore need to be established, nurtured, and maintained. For effective implementation, we need to ensure the following:-

- All universities should liaise with the relevant industries to receive industrial knowledge to augment the classroom lectures.
- University lecturers should be motivated towards attending local and international workshops on the latest IT innovations with the intension of transferring same knowledge to their students.
- University should encourage students towards registration for certifications in IT.
- To consider university education as a "preparation for life" and therefore should cover a wider spectrum of courses that will be relevant in industry.
- The existing gap can also be corrected by reviewing the whole of the university curriculum (Information Technology), and by preparing lecturers/instructors in line with the new curriculum because implementation is another challenge when it comes to curriculum review.
- Retraining of the existing teaching staff and administrators and redefining/restructuring teacher preparation programs in keeping with the new requirements in IT.
- IT education is the minimum requirement for survival in today's society and should therefore be open for universal access.
- Information Technology (IT) should be integrated in all the subjects of the curriculum at primary, secondary and tertiary levels.
- The colleges of education and the University will have to change the way they prepare teachers in keeping with the new requirements. Both in-service and pre-service programs have to be developed to serve this purpose.
- The preparation of teachers has to start as soon as possible as this is a long term process.
- Also, we have to properly fund our universities, quantitatively or qualitatively so that our citizenry, including our labor force, may be sufficiently empowered with appropriate knowledge of $21^{st}$ century skills and attitude for effective participation in a very competitive global society of IT.
- Implement faculty improvement programs to upgrade their caliber and learn new technologies based on suggestions of leading software industrialists.
- Focus on industrial driven needs that will enhance the chance of university graduates rather than laying emphasis much on the basis; that is, the underlying principle of computing.
- The gap between Industrial based applications and university curriculum can also be bridged if the curriculum can be structured in ways that will concur with the industrial applications
- Also, this can be achieved if computer science courses related to application development such as programming can be taken by professionals in such field that are currently working in industry.

This will require adjustments to the curriculum and format of universities education. It will also require universities to be more open to constructive engagement with employers of labor itself, as well as encouraging them to share their hands on experience with, and inspire university students while they are still in school.

On the hand, universities need significant funding improvements for research, learning and related intellectual activities, intellectual freedom, the scope to think and interact with academics in many locations and circumstances, articulate and operate semi-autonomously such that those who provide the funding should not therefore believe that all things related to their funding must be done their way at all time.

From the foregoing, it is obvious that bridging the skill gaps is not merely improving students' competence in core fields of IT. Education with relevant syllabuses and training in specific areas play crucial roles in achieving rapid changes in updating technical and engineering skills especially in making relevant the degree in computing and IT skill demand in our industries.